\documentclass[twocolumn,english,preprintnumbers,amsmath,amssymb,pre]{revtex4}

\usepackage{graphicx}

\begin{document}

\title{Water at negative pressure: Nuclear quantum effects}
\author{Carlos P. Herrero}
\author{Rafael Ram\'irez}
\affiliation{Instituto de Ciencia de Materiales de Madrid,
         Consejo Superior de Investigaciones Cient\'ificas (CSIC),
         Campus de Cantoblanco, 28049 Madrid, Spain }
\date{\today}

\begin{abstract}
Various condensed phases of water, spanning from the liquid state to 
multiple ice phases, have been systematically investigated under extreme 
conditions of pressure and temperature to delineate their stability 
boundaries. This study focuses on probing the mechanical stability of 
liquid water through path-integral molecular dynamics simulations, 
employing the q-TIP4P/F potential to model interatomic interactions 
in flexible water molecules.
Temperature and pressure conditions ranging from 250 to 375~K and 
$-0.3$ to 1~GPa, respectively, are considered. This comprehensive approach 
enables a thorough exploration of nuclear quantum effects on various 
physical properties of water through direct comparisons with classical 
molecular dynamics results employing the same potential model. 
Key properties such as molar volume, intramolecular bond length, 
H--O--H angle, internal and kinetic energy are analyzed, 
with a specific focus on the effect of tensile stress.
Particular attention is devoted to the liquid-gas spinodal pressure, 
representing the limit of mechanical stability for the liquid phase, 
at several temperatures. The quantum simulations reveal a spinodal 
pressure for water of $-286$ and $-236$ MPa at temperatures of 250 
and 300~K, respectively. At these temperatures, the discernible shifts 
induced by nuclear quantum motion are quantified at 15 and 10~MPa, 
respectively. These findings contribute valuable insights into 
the interplay of quantum effects on the stability of liquid water under 
diverse thermodynamic conditions.  \\

\end{abstract}

\maketitle

\section{Introduction}

In recent decades, the experimentally accessible domain within the phase 
diagrams of diverse substances has considerably expanded. This enlargement
has provided a more profound comprehension of condensed phases subjected 
to extreme conditions of temperature and pressure, as documented by various 
studies \cite{sc-mu03,sc-ma18}. Consequently, there has been an in-depth 
exploration of how different properties of condensed matter respond to 
hydrostatic pressure.
This scrutiny extends to a growing interest in tensile stress, a factor 
that holds the potential to enhance our understanding of the metastability 
boundaries inherent in various phases. 
\cite{sc-da10,sc-iy14,sc-ni19,sc-im08c}. 
This contributes to the refinement of our knowledge about the behavior 
of condensed matter under extreme conditions and provides insights into 
the attractive region of intermolecular forces.

Over the years, the high-pressure region of the phase diagram for water 
has garnered significant interest within the realms of condensed matter 
physics, chemistry, and planetary sciences, as evidenced by a body of 
research \cite{w-ei69,w-pe99,w-sa04,w-sa19,w-zh21}. 
This attention is not only attributed to the intrinsic importance of water,
but is also motivated by the multitude of ice polymorphs discovered under 
varying conditions of temperature ($T$) and pressure ($P$).
To date, investigations into the phase diagram of water have been 
extended to temperatures up to $T \sim 1000$~K and pressures 
reaching several hundreds of GPa \cite{w-du10}. 

The examination of condensed matter behavior under tensile pressure has 
predominantly focused on liquids
\cite{sc-so92,bo94b,sc-je03,sc-im07,sc-im08,sc-da10}. These investigations 
delve into the mechanical stability limits and the occurrence of cavitation 
in proximity to their spinodal lines.
In recent years, the experimentally accessible domain of hydrostatic 
(or quasi-hydrostatic) tensile pressure has significantly enlarged. 
This expansion has not only widened the scope of study, but has also 
deepened our understanding of physical properties under conditions that 
pose challenges in laboratory settings 
\cite{sc-da10,sc-he87,sc-gr88,sc-mo00,sc-du02,sc-ve22}. 

Liquid water under tensile pressure has undergone extensive scrutiny 
in recent years, with both theoretical investigations, 
primarily through simulations 
\cite{w-po92,w-ab11,w-bi21,w-ga07,w-im13,w-ne01,w-pa16,w-se17,w-si18,w-st02b}, 
and experimental studies 
\cite{sc-da10,w-az13,w-pa14,w-al93,w-qi16,w-st16,w-ho17}.
Numerous inquiries have focused on identifying thermodynamic and mechanical 
anomalies within stretched water. Variables such as density, compressibility, 
and sound velocity have been subject to exploration in various works 
\cite{w-al17,w-ca12,w-pa14,w-bi17,w-ho17,w-ca19,w-za20}. Noteworthy attention 
has been given to investigating the phenomenon of water cavitation under 
negative pressure, wherein the metastable liquid undergoes breakdown through 
the nucleation of vapor bubbles \cite{w-az13,w-go14,w-me16,w-ma21,w-sa23,w-la23}. 
Within this context, several studies have delved into the liquid-gas spinodal 
line of water \cite{w-po92,w-ne01,w-ga07,w-ca19,w-du20,w-im08d}.

In this paper, we focus on the properties of water under negative 
pressure. Our main goal is to gain insight into the influence of
nuclear quantum motion on such properties, particularly
on the liquid-gas spinodal line. To achieve this, we have
carried out extensive path-integral molecular dynamics (PIMD)
simulations using a reliable interatomic potential (q-TIP4P/F),
suitable for this type of simulations.
We find that the spinodal pressure changes by 15 and 10~MPa at
$T = 250$ and 300~K, respectively, as compared with the results of
classical molecular dynamics simulations.
Quantum effects are also analyzed for the molar
volume, structure of water molecules, as well as internal and 
kinetic energy of the liquid in the region of negative pressure.
Similar techniques based on atomistic simulations have been
employed earlier to study the mechanical stability of 
solids \cite{he03b,w-ma19,sc-he23}.

The paper is structured as follows: In Sec.~II, we detail the 
computational method utilized in our calculations. Sec.~III is 
dedicated to the discussion of kinetic and internal energy as 
functions of pressure and temperature. Structural properties, 
including the O--H bond length and H--O--H angle, are examined 
in Sec.~IV, while Sec.~V delves into the molar volume.
The liquid-gas spinodal instability is explored in Sec.~VI, and
Sec.~VII summarizes the main results.

\section{Computational Method}

In this study, we employ the PIMD computational technique to investigate 
the properties of water across a range of temperatures and pressures, 
encompassing both compression ($P > 0$) and tension ($P < 0$) conditions. 
The foundation of this approach lies in an isomorphism established 
between the actual quantum system and an artificial classical counterpart, 
which emerges through the discretization of the quantum density matrix 
along cyclic paths \cite{fe72,kl90}.
This isomorphism is practically realized by substituting each quantum 
particle with a ring polymer composed of $N_{\rm Tr}$ (Trotter number) 
classical pseudo-particles. These pseudo-particles are interconnected 
by harmonic springs with force constants that vary with mass and temperature. 
The isomorphism is exact for $N_{\rm Tr} \to \infty$, but in 
practical applications, the choice of $N_{\rm Tr}$ involves a trade-off 
between accuracy (increasing $N_{\rm Tr}$) and computational feasibility 
(reducing $N_{\rm Tr}$).
Further details on this simulation method can be found 
in previous works \cite{gi88,ce95,he14}.

The dynamics observed in PIMD simulations are inherently artificial, as 
they do not faithfully capture the genuine quantum dynamics of atomic 
nuclei. However, despite this artificiality, PIMD simulations 
serve as a practical and potent approach for effectively exploring 
the many-body configuration space. This capability enables 
the generation of precise data regarding the equilibrium properties 
of the actual quantum system.
An alternative method for computing equilibrium properties relies on 
Monte Carlo sampling. Nonetheless, this approach demands greater 
computational resources, notably in terms of CPU time, compared to 
the PIMD method, which holds an advantage in this context 
due to its enhanced parallelization potential. This feature is crucial 
for optimizing the efficiency of operations on modern 
computer architectures.

For our water simulations, we employed the point charge, flexible 
q-TIP4P/F potential model to describe interatomic interactions. 
This model has been demonstrated to be particularly suitable for 
analyzing nuclear quantum effects in the structural, dynamical, and 
thermodynamic properties of water \cite{w-ha09}. Notably, it has been 
applied to investigate various states of water, including the liquid 
phase \cite{w-ha09,w-ha09b,w-ra11,w-ha17,w-el21}, 
ice \cite{w-ra10,w-he11,w-ha11b,w-he11b,w-ra19,w-he15}, 
clusters \cite{w-go10}, and the phase diagram \cite{w-ra13}. 
Furthermore, the same model has been recently utilized to examine 
a liquid-liquid phase transition in water \cite{w-el22}.

Other simulations investigating condensed phases of water have relied 
on empirical potentials that treat H$_2$O molecules as rigid entities 
\cite{w-he05,w-mi05,w-he06b}. While this approach can offer computational 
efficiency and produce satisfactory results for various properties of 
the liquid state and various ice polymorphs, 
it overlooks the impact of molecular flexibility 
on the dynamics and structure of different phases \cite{w-ha09}.
For instance, employing flexible H$_2$O molecules allows for the exploration 
of correlations between intramolecular O--H bond distances ($d_{\rm O-H}$) 
and the geometry of intermolecular H bonds, a phenomenon extensively studied 
in ice \cite{w-ho72,w-ny06,w-pa12}. Additionally, incorporating anharmonic 
stretches in the q-TIP4P/F potential enables the examination of variations 
in the $d_{\rm O-H}$ distance as a function of both temperature and pressure. 
This approach further facilitates the discernment of differences between 
data derived from classical and quantum simulations.

\begin{figure}
\vspace{-6mm}
\includegraphics[width=7cm]{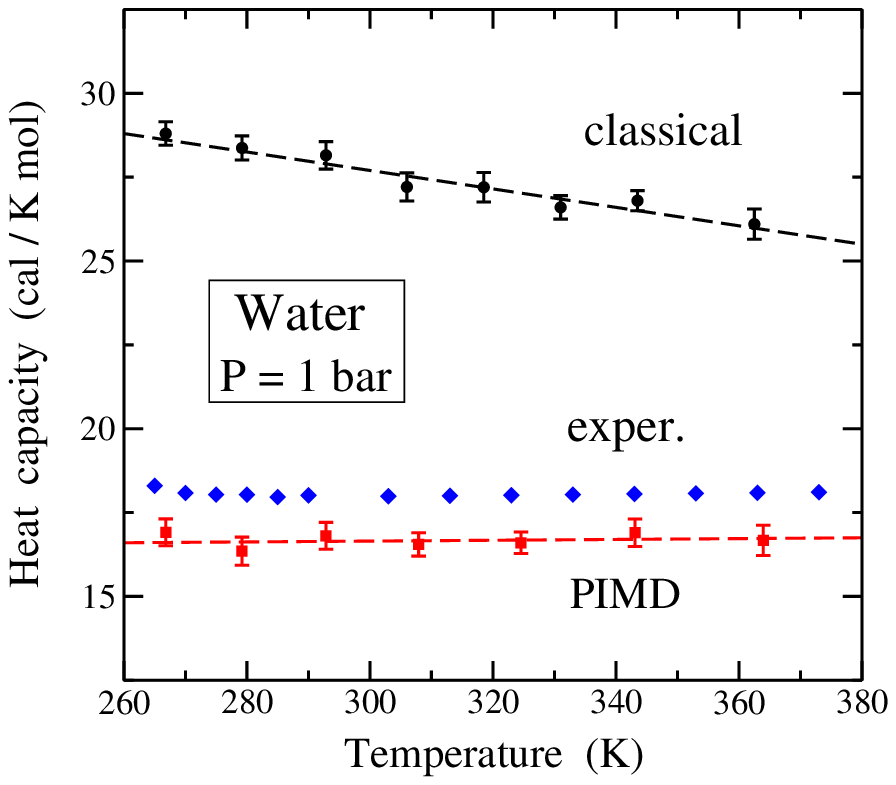}
\vspace{-5mm}
\caption{Heat capacity of water as a function of temperature,
as derived from classical MD (circles) and PIMD simulations
(squares). Diamonds indicate experimental values given by
Angell {\em et al.} \cite{w-an82}, denoted as 'exper.'.
Dashed lines are guides to the eye.
}
\label{f1}
\end{figure}

Most of the water simulations discussed herein were conducted within 
the isothermal-isobaric $NPT$ ensemble. This choice facilitated 
the determination of the equilibrium volume of the liquid under 
specified pressure and temperature conditions. The PIMD simulations 
in this statistical ensemble were executed using algorithms detailed 
in the literature \cite{ma96,tu98,ma99,tu02}.
Staging variables were employed to define 
the bead coordinates, and a constant temperature was maintained 
by coupling chains of four Nos\'e-Hoover thermostats to each staging 
variable. Additionally, a chain incorporating four barostats 
was connected to the volume to uphold the desired pressure, as 
described by Tuckerman and Hughes \cite{tu98}. 
The pressure was calculated using a virial expression 
suitable for PIMD simulations \cite{he14}. 
Near the liquid-gas spinodal (negative) pressure, $P_s$, we also 
conducted PIMD simulations in the canonical ($NVT$) ensemble. 
This approach allowed for extended simulations of metastable liquid 
water in a region of the phase diagram where $NPT$ simulations tend 
to become rapidly unstable.

For our simulations, we utilized cubic cells containing 300 water molecules 
with periodic boundary conditions. Configuration space sampling was 
conducted over a temperature range of 250 K to 375~K and pressures 
ranging from $-0.3$ to 1~GPa. The Trotter number
was chosen to be proportional to the inverse temperature, specifically 
satisfying $N_{\rm Tr} \, T = 6000$~K.
Electrostatic interactions in the q-TIP4P/F potential were calculated  
by the Ewald method.
The equations of motion were integrated using the reversible reference 
system propagator algorithm (RESPA), allowing for the use of different 
time steps for the integration of slow and fast dynamical 
variables \cite{ma96}. Interatomic forces were computed with a time 
interval of $\Delta t = 0.2$~fs, which demonstrated acceptable 
convergence for the considered variables.
In a typical simulation run at a given temperature $T$ and pressure $P$, 
we performed $10^5$ PIMD steps for system equilibration and subsequently 
conducted $8 \times 10^6$ steps for computing average properties. 
For $NVT$ simulations near the liquid-gas spinodal, runs included 
$2 \times 10^7$ steps to enhance statistics in the region of 
negative pressure.
Other technical details about the simulations presented here were
previously described in \cite{w-he11,w-he11b}.

To evaluate the extent of quantum effects in the outcomes of PIMD 
simulations, we carried out classical molecular dynamics (MD) simulations 
of water employing the same interatomic potential q-TIP4P/F. 
In our context, this is akin to setting the Trotter number 
$N_{\rm Tr}$ equal to 1.

\section{Energy}

\begin{figure}
\vspace{-6mm}
\includegraphics[width=7cm]{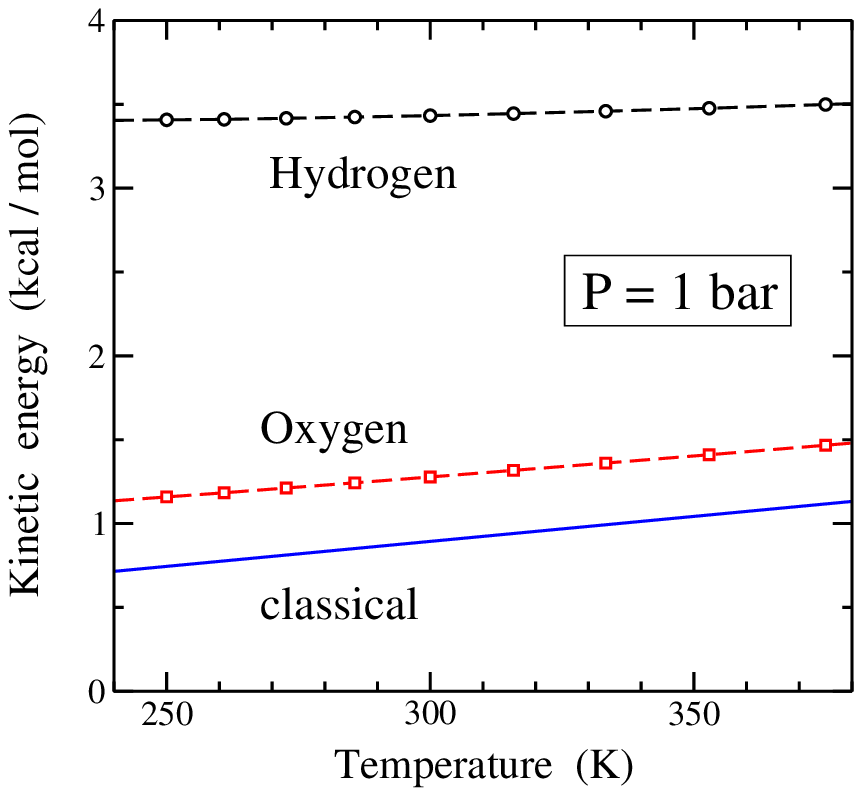}
\vspace{-5mm}
\caption{Kinetic energy as a function of temperature for
$P$ = 1 bar, derived from PIMD simulations for hydrogen (circles)
and oxygen (squares) in water. Error bars are smaller than the
symbol size.  The solid line indicates the classical value:
$E_k^{\rm cl} = 3 R T / 2$.
Dashed lines are guides to the eye.
}
\label{f2}
\end{figure}

In this section, we delve into the internal energy of water, 
derived from our PIMD simulations, with particular emphasis on the 
kinetic energy component. As an additional check of the potential 
model, we computed the heat capacity, $c_p$, by numerically 
differentiating the enthalpy $H = E + P V$, i.e., $c_p = d H /d T$. 
Fig.~1 illustrates the temperature-dependent behavior of $c_p$ at 
a constant pressure of 1~bar, featuring results obtained from both 
classical MD simulations (depicted by circles) and PIMD simulations 
(shown as squares).
Within this figure, diamonds symbolize the experimental 
heat capacity of water at several temperatures, given by 
Angell {\em et al.} \cite{w-an82}.
The classical MD results overestimate the actual 
heat capacity, exhibiting a decreasing trend with rising temperature 
throughout the displayed temperature range in Fig.~1. This trend 
reduces the disparity with the quantum data. On the other hand, 
the outcomes of PIMD simulations align more closely with the experimental 
values, although they fall slightly below the latter. This difference 
exceeds the error bar associated with the simulation data.

We note that using alternative effective potentials for classical 
simulations can yield better agreement with known properties of water,
as compared to the q-TIP4P/F potential in our classical 
simulations \cite{w-sh16}. In this line, one could contrast results 
of classical and PIMD simulations utilizing different potentials,
but this can hinder a direct assessment of quantum effects. 
This is because such effects could be confounded by the inherent 
distinctions between the interatomic potentials themselves. 
Our goal is mainly centered on quantifying the magnitude of
nuclear quantum effects, assuming a reliable interatomic potential, 
rather than discerning between the quality of various potentials.

Path integral simulations offer distinct estimators for the potential 
($E_p$) and kinetic energy ($E_k$) of a system. In the domain 
of molecular and condensed-matter physics, the kinetic energy of an 
atomic nucleus depends not only on temperature but also on its mass 
and spatial delocalization.
Classical physics posits that each degree of freedom contributes to 
the kinetic energy per mole in an amount proportional to the temperature, 
specifically $R \, T / 2$ (where $R$ is the gas constant), aligning with 
the equipartition principle. However, the quantum kinetic energy 
is linked to the potential landscape surrounding the particle 
under consideration. Path integral simulations emerge as an apt technique 
for computing $E_k$ of quantum particles, providing insights into 
the dispersion of quantum paths.
In particular, the radius-of-gyration of the cyclic paths becomes an
important factor, as a larger radius-of-gyration corresponds to greater 
quantum delocalization, resulting in a reduction of $E_k$ \cite{gi88,w-he11}.
We have undertaken the calculation of $E_k$ in water using the virial 
estimator, acknowledged for its statistical precision \cite{he82,tu98}, 
yielding error bars smaller than those associated with the potential energy, 
especially at elevated temperatures \cite{w-ra11,he22}. The kinetic energy 
of hydrogen in water at atmospheric pressure was previously studied in detail 
through PIMD simulations, especially in connection 
with deep inelastic neutron-scattering experiments \cite{w-ra11}.

In Fig.~2, we present the kinetic energy per mole of oxygen (depicted 
as squares) and hydrogen (depicted as circles) as a function of 
temperature, calculated from our PIMD simulations performed at 
$P = 1$~bar. $E_k^{\rm H}$ for hydrogen atoms is clearly
larger than for oxygen atoms, $E_k^{\rm O}$, as expected for the smaller
mass of the former.
 Within the temperature range illustrated in Fig.~2, 
the ratio $E_k^{\rm H} / E_k^{\rm O}$ decreases from 3.0 to 2.4.
The cumulative kinetic energy of water,
$E_k = 2 E_k^{\rm H} + E_k^{\rm O}$, exhibits an ascent from 
7.93 kcal/mol at 250~K to 8.46 at 375~K, with $E_k =$~8.14 kcal/mol 
at 300~K. The solid line in Fig.~2 represents the classical value of 
$E_k^{\rm O}$ or $E_k^{\rm H}$, i.e. $E_k^{\rm cl} = 3 R \, T / 2$.
In the temperature range depicted in this figure, the quantum kinetic 
energy of oxygen gradually converges toward the classical value as 
$T$ increases. However, the former still remains noticeably greater 
than the latter at temperatures around 400~K.

\begin{figure}
\vspace{-6mm}
\includegraphics[width=7cm]{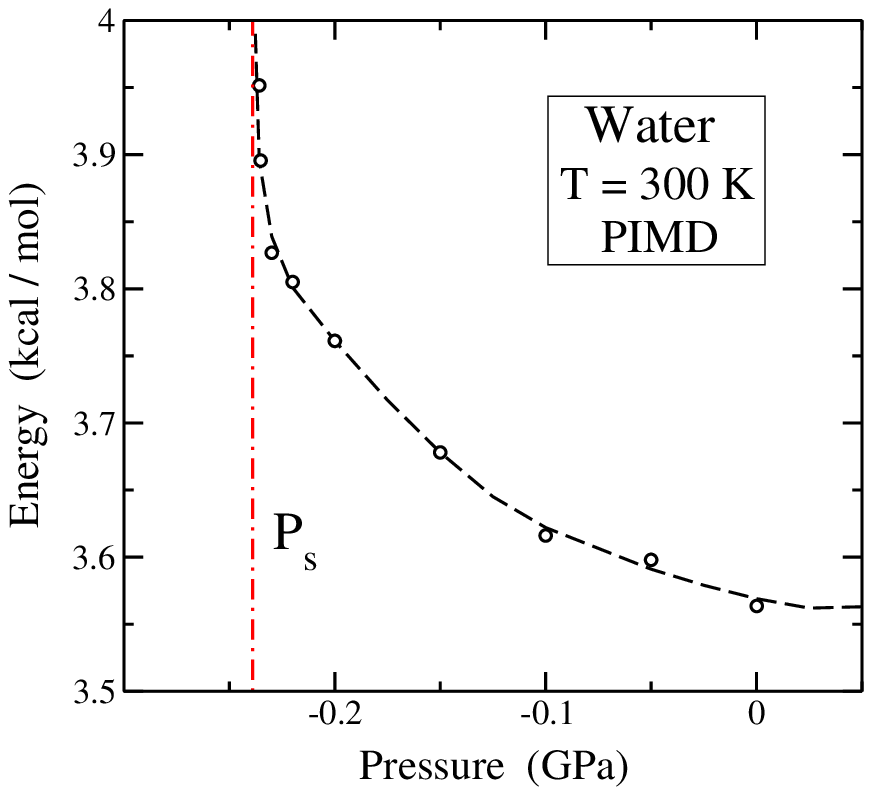}
\vspace{-5mm}
\caption{Pressure dependence of the energy for $T$ = 300 K
(circles).  Symbols represent results of PIMD simulations.
Error bars are on the order of the symbol size.
The dashed line is a guide to the eye.
A vertical dashed-dotted line indicates the pressure of
mechanical instability at 300~K.
}
\label{f3}
\end{figure}

We now explore the pressure-dependent behavior of the internal and 
kinetic energy of water. In Fig.~3, we display the internal energy,
$E = E_p + E_k$, as a function of pressure, 
derived from our PIMD simulations utilizing the q-TIP4P/F interatomic 
potential for $P \leq 0$. The energy scale in this plot aligns with 
the original parameterization of the potential model \cite{w-ha09}.
With an increasing tensile pressure, the internal energy steadily 
rises until $P$ reaches values around $-0.22$~GPa, where a prominent 
cusp appears. This cusp signifies the proximity to the mechanical 
instability of the liquid. The elevation of energy under increasing 
tensile pressure primarily results from the growth of the system's 
potential energy, which distinctly dominates over the decrease in 
kinetic energy, as illustrated in Fig.~4.
The vertical dashed-dotted line in Fig.~3 represents the critical 
pressure $P_s = -0.236$~GPa, denoting the limit of mechanical 
stability at $T = 300$~K, as explained in Sec.~VI.

\begin{figure}
\vspace{-6mm}
\includegraphics[width=7cm]{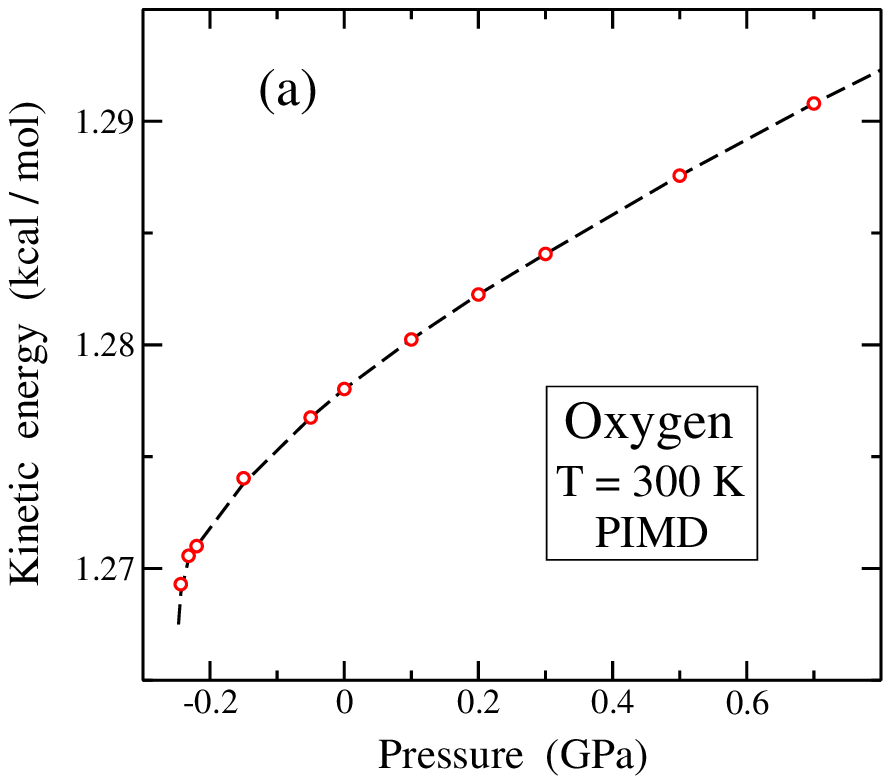}
\includegraphics[width=7cm]{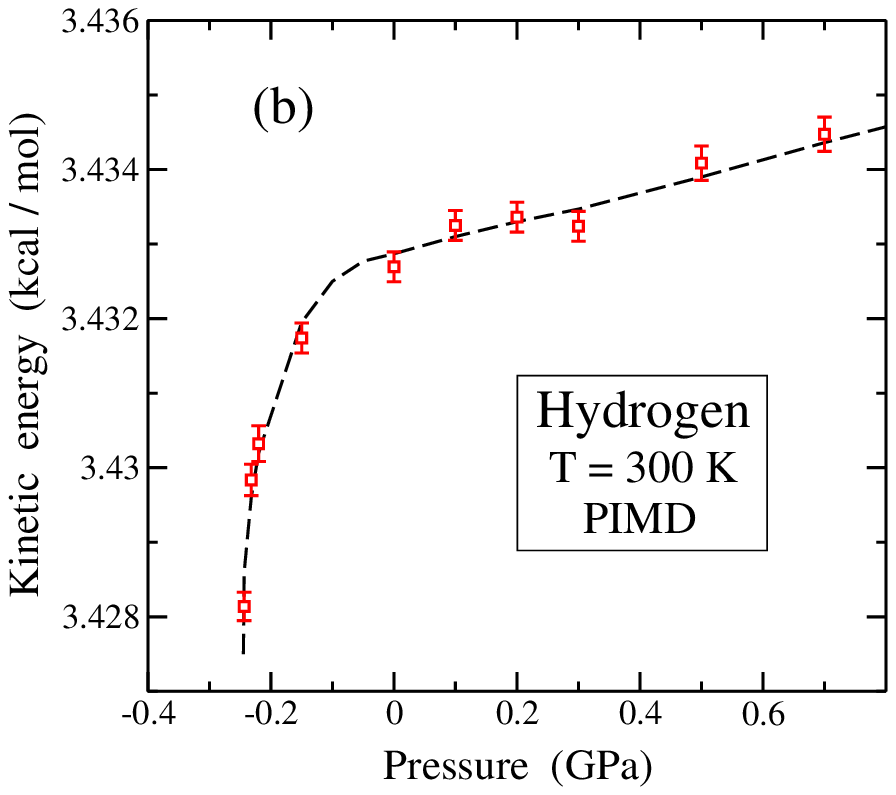}
\vspace{-5mm}
\caption{Kinetic energy as a function of pressure for
$T$ = 300~K, calculated from PIMD simulations.
(a) Oxygen (open circles); (b) hydrogen (open squares).
Error bars in (a) are on the order of the symbol size.
The dashed lines are guides to the eye.
}
\label{f4}
\end{figure}

In Fig.~4, we show the kinetic energy evolution of (a) oxygen 
and (b) hydrogen atoms at 300~K as a function of $P$, 
encompassing compressive and tensile stress conditions. 
For both atomic species, there is a discernible augmentation in $E_k$
as compressive stress intensifies, indicative of $d E_k / d P > 0$.
This positive derivative persists under tensile stress ($P < 0$); 
however, its magnitude undergoes a rapid surge around $P = -0.2$~GPa.
This pronounced increase is particularly conspicuous for hydrogen, 
especially in close proximity to the mechanical instability of water.

For a given temperature, an elevation in kinetic energy correlates 
with a reduction in the dispersion of quantum paths, as indicated by 
the decrease in the radius-of-gyration associated with the considered 
atomic nucleus. Consequently, both O and H paths contract under 
increasing compressive pressure and expand under tensile stress.
The upward trend exhibited by the kinetic energy for $P > 0$ aligns with 
the typical pattern of heightened vibrational frequencies 
observed with increasing pressure in condensed matter. 
However, it is noteworthy that the change in $E_k^{\rm O}$ and 
$E_k^{\rm H}$ throughout the entire pressure range 
depicted in Fig.~4 is considerably smaller than the growth observed in 
Fig.~2 when temperature increases from 250 to 375~K, specifically 
0.31 and 0.09 kcal/mol for oxygen and hydrogen, respectively.

\section{Molecular geometry}

In this section we present results for the molecular geometry of water,
which can shed light on the structural changes suffered by the
liquid when temperature or pressure are modified.
We concentrate on the intramolecular O--H bond length and the H--O--H angle.

\begin{figure}
\vspace{-6mm}
\includegraphics[width=7cm]{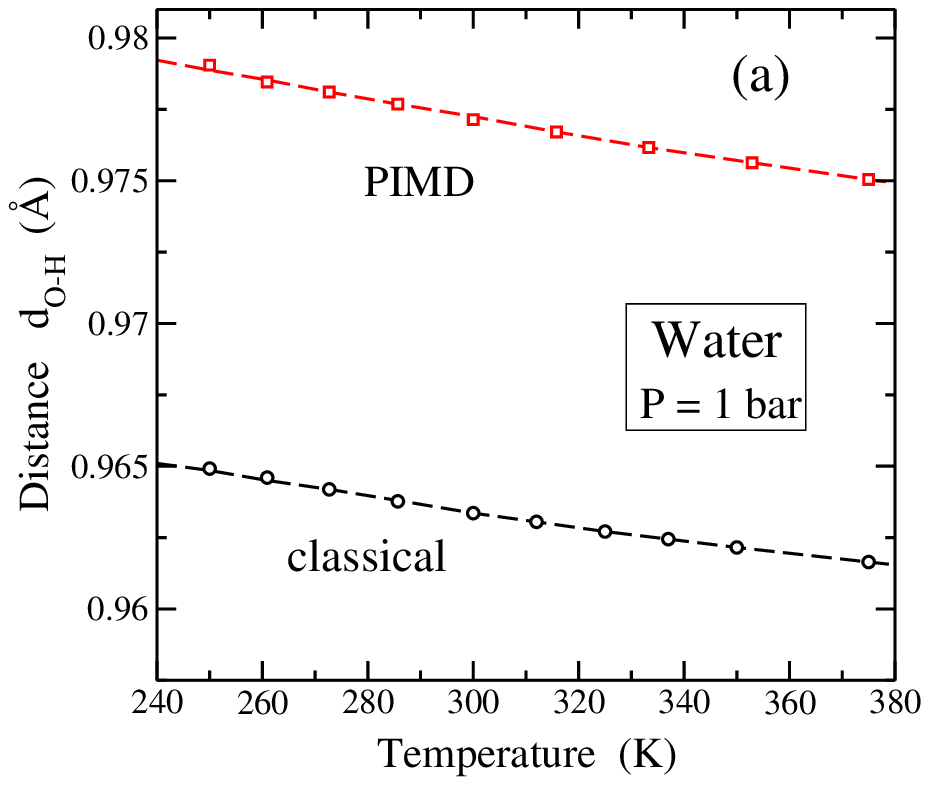}
\includegraphics[width=7cm]{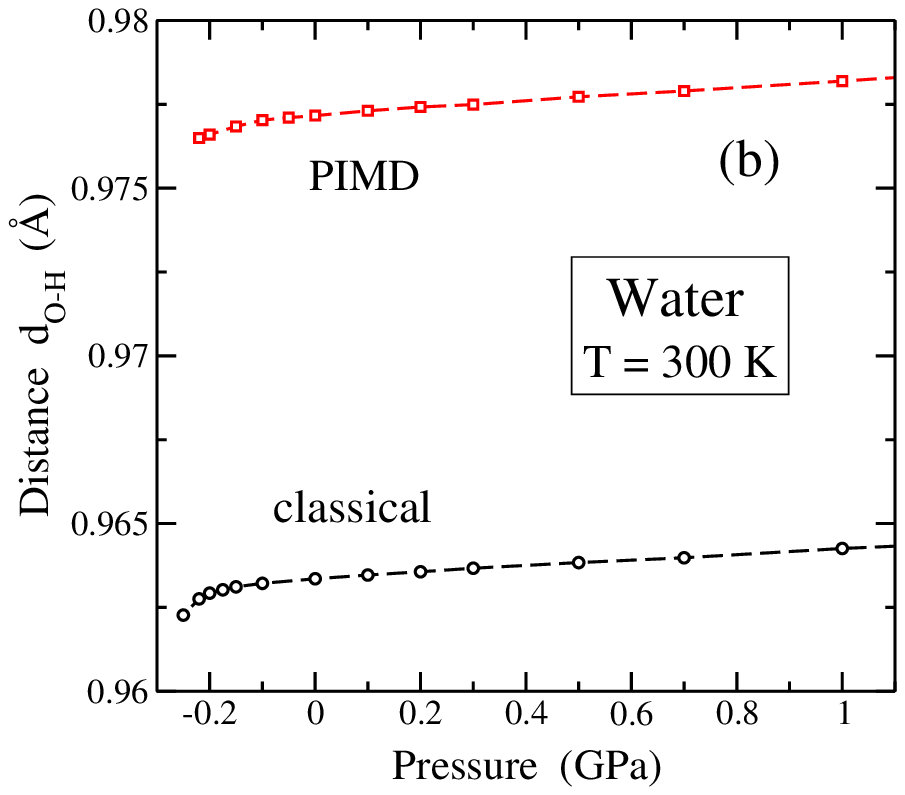}
\vspace{-5mm}
\caption{(a) Intramolecular O--H distance, $d_{\rm O-H}$, vs temperature
for $P = 1$~bar, as derived from simulations with the q-TIP4P/F potential.
(b) $d_{\rm O-H}$ as a function of pressure for $T = 300$~K.
In both panels, circles and squares represent
results of classical MD and PIMD simulations, respectively.
Error bars are in the order of the symbol size.
Lines are guides to the eye.
}
\label{f5}
\end{figure}

\subsection{O--H bond}

Fig.~5 illustrates the temperature and pressure dependencies of the bond 
distance $d_{\rm O-H}$ in water, as determined through classical and 
quantum simulations using the q-TIP4P/F potential model. In Fig.~5(a) 
the mean distance $d_{\rm O-H}$ is presented as a function of temperature 
at $P$ = 1~bar. Simulation results are denoted by 
open symbols, with circles representing classical MD and squares 
corresponding to PIMD. Classical and quantum simulations 
exhibit a similar temperature dependence, as depicted by the 
trends observed in results for both cases.

In Fig.~5(a) it is clear that the mean distance $d_{\rm O-H}$
resulting from quantum simulations surpasses that obtained in classical 
simulations. Specifically, for water the disparity between the two 
datasets is 0.014~\AA\ (1.4\% of the bond length), and this 
difference marginally diminishes by less than $10^{-3}$~\AA\ within 
the considered temperature range. At $T$ = 300~K, the rate of 
change of $d_{\rm O-H}$ with respect to temperature is 
$\partial d_{\rm O-H} / \partial T = -2.7 \times 10^{-5}$ \AA/K 
for classical results and $-3.1 \times 10^{-5}$ \AA/K for the quantum data. 
Notably, the bond expansion induced by nuclear quantum motion mirrors 
findings observed in analogous PIMD simulations for ice Ih \cite{w-he11b}.

We highlight a noteworthy observation in our findings: the decrease in 
$d_{\rm O-H}$ as temperature rises. At first glance, this may appear 
counterintuitive, especially considering the conventional understanding 
of thermal expansion in atomic bonds. The counterintuitive nature of 
this trend becomes more apparent when considering the expected 
expansion associated with an increase in volume.
However, our results unveil an intriguing and seemingly anomalous 
behavior that aligns with observations in various water phases, where 
the covalent O--H bond contracts with rising temperature 
($\partial d_{\rm O-H} / d T < 0$). This behavior arises from 
a delicate interplay between the general tendency of bond distances 
to increase with temperature and the opposing effect caused by 
the fortification of the intramolecular O--H bond.
This fortification is linked to a weakening of intermolecular H bonds, 
resulting from a larger mean O--O distance and the consequential expansion 
of volume. In essence, our findings agree with the local 
intra-intermolecular geometric correlation identified in both liquid 
and solid water. This correlation relates the intramolecular O--H bond 
length to the corresponding H-bond geometry \cite{w-ho72,w-ny06}.
Similar effects on $d_{\rm O-H}$ have been previously documented in 
PIMD simulations of ice VII using the q-TIP4P/F potential 
model \cite{w-he15}. 

Considering the aforementioned arguments, we anticipate an increase 
in the bond distance $d_{\rm O-H}$ under increasing compressive 
pressure. In this scenario, the reduction in volume leads to 
a contraction of intermolecular H bonds, consequently causing 
the expansion (weakening) of intramolecular bonds. This relationship 
is illustrated in Fig.~5(b), where $d_{\rm O-H}$ is plotted against 
hydrostatic pressure $P$.
The figure reveals a consistent trend across classical and quantum 
datasets, showing that $d_{\rm O-H}$ expands with increasing 
compression and contracts with growing tensile pressure. 
The difference between the two datasets remains constant within 
the pressure range depicted in Fig.~5(b), totaling 0.014~\AA. 
For $P > 0$, the change in $d_{\rm O-H}$ is characterized by a slope 
$\partial d_{\rm O-H} / \partial P = 1.0 \times 10^{-3}$ \AA/GPa,
observed consistently in both classical and quantum results.
Conversely, for $P < 0$ a reduction in the bond distance is observed, 
accelerating as the system approaches mechanical instability at 
$P \sim -0.24$~GPa.  At the scale of Fig.~5(b), this 
reduction does not exhibit as sharp a transition as observed in 
the kinetic energy presented in Fig.~4.

We observe that variations in the interatomic distance $d_{\rm O-H}$, 
depicted in Figs. 5(a) and 5(b), appear relatively minor. However, 
they are comparable to discrepancies observed among intramolecular distances 
in different ice phases \cite{w-he15}. Moreover, these changes may hold 
significance in elucidating variations in vibrational frequencies within 
water under varying external conditions.

\begin{figure}
\vspace{-6mm}
\includegraphics[width=7cm]{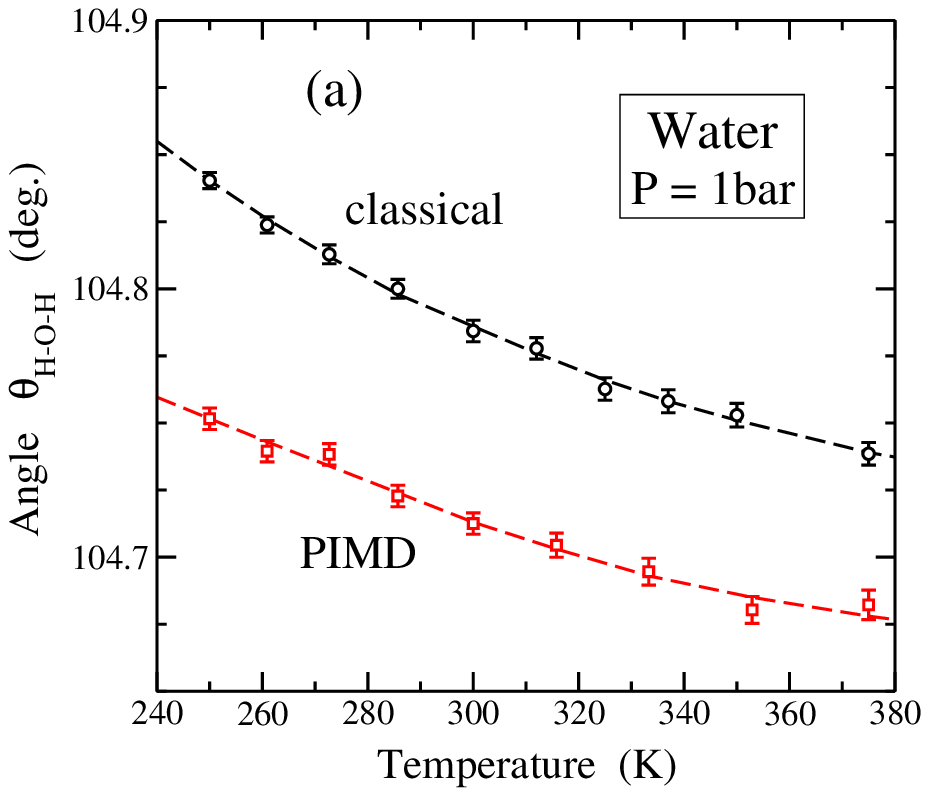}
\includegraphics[width=7cm]{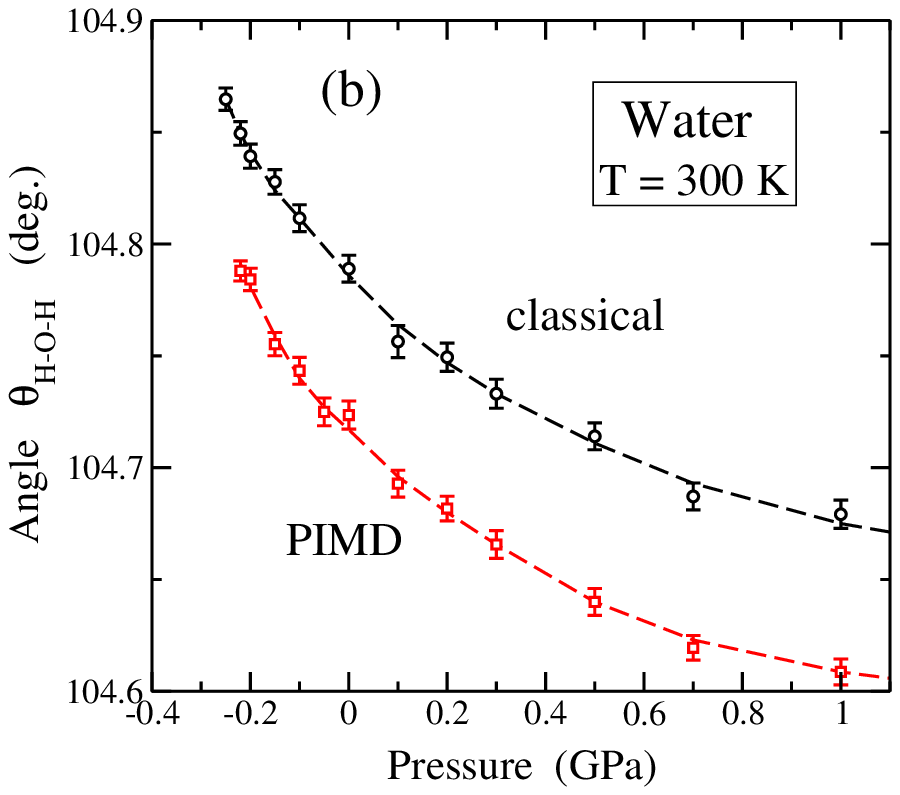}
\vspace{-5mm}
\caption{Mean angle $\theta_{\rm H-O-H}$ obtained from simulations with
the q-TIP4P/F potential. (a) Temperature dependence for $P$ = 1 bar;
(b) Pressure dependence for $T = 300$~K.
Circles and squares indicate results of
classical MD and PIMD simulations, respectively.
Lines are guides to the eye.
}
\label{f6}
\end{figure}

\subsection{H--O--H angle}

In Fig.~6(a), we present a comprehensive analysis of the 
temperature-dependent behavior of 
the mean H--O--H angle, denoted as $\theta_{\rm H-O-H}$, 
in water molecules, carried out using classical MD 
(represented by circles) and PIMD simulations (indicated by squares). 
Our results at 300~K are in agreement with the findings from 
Refs.~\cite{w-ha09,w-ra10}, where the q-TIP4P/F interatomic potential 
was employed.

Upon examination of both classical and quantum datasets in Fig.~6(a), 
a consistent pattern emerges: 
an observable reduction in the angle $\theta_{\rm H-O-H}$
as the temperature is elevated. This fact holds significance 
for understanding the thermal dynamics of water molecules. 
A decrease in $\theta_{\rm H-O-H}$ for rising $T$ was also found 
by Gu~{\em et al.} from classical MD simulations using
the MCYL potential \cite{w-gu96}.  Furthermore, our investigation 
reveals that nuclear quantum motion induces a subtle but 
discernible diminution in the intramolecular angle, approximately 
0.1 deg., with this effect becoming less pronounced at higher 
temperatures.  This interplay between temperature and quantum 
effects sheds light on the dynamical aspects of the water molecule 
behavior in the liquid phase.
At $T = 300$~K, we find $\partial \theta_{\rm H-O-H} / \partial T$
= $-8.6 \times 10^{-4}$ deg./K and $-6.1 \times 10^{-4}$ deg./K from
our classical and quantum results, respectively. 
The change with the temperature is larger for the classical model,
and both sets of results approach to each other as $T$ rises.

Expanding our analysis to include the effect of pressure, Fig.~6(b) 
showcases the mean angle $\theta_{\rm H-O-H}$ under compressive and
tensile pressure conditions. 
Both classical (circles) and quantum (squares) simulations 
at $T$ = 300~K present a decrease in the mean angle for rising $P$,
i.e., $\partial  \theta_{\rm H-O-H} / \partial P < 0$, in the whole 
range of pressure under consideration. 
At atmospheric pressure, we obtain from our simulations a
pressure derivative $\partial \theta_{\rm H-O-H} / \partial P$
= --0.26(1) deg./GPa for both the classical and quantum data,
which coincide within the precision of our numerical procedure.
This slope clearly becomes less negative as compressive pressure
is raised.  On the contrary,
when subjected to increasing tensile stress within 
the mechanical stability region of the liquid phase, the mean angle 
$\theta_{\rm H-O-H}$ exhibits growth. 
This response to compressive and tensile pressures 
provides insight into the structural adaptability of water 
molecules under different conditions.

A noteworthy observation is the amplification of the negative 
slope in the vicinity of the stability limit ($P \sim -0.24$~GPa).
This indicates a heightened sensitivity of the H--O--H 
angle to pressure variations, especially in the proximity of the 
spinodal pressure $P_s$ corresponding to each approach (classical 
or quantum). 

Water molecules within the crystal structures of diverse ice phases 
exhibit a notable range in bond angles, spanning from 96.4 to 
112.8 deg. For isolated water molecules, detailed quantum-mechanical 
calculations \cite{w-mi20} have unveiled a discernible 
negative correlation between the distance $d_{\rm O-H}$ and the angle 
$\theta_{\rm H-O-H}$. This correlation aligns with 
the observed pressure dependence in liquid water, illustrated in 
Figs.~5(b) and 6(b).
In contrast, from the temperature dependence analysis of both structural
variables we find a positive correlation, as portrayed in Figs.~5(a)
and 6(a) (mean bond length and angle decrease for rising $T$). 
This can be related to an expanding dispersion in the actual bond 
and angle distributions as temperature increases \cite{w-el21}.

To conclude this section, it is noteworthy that changes in the angle 
$\theta_{\rm H-O-H}$, as depicted in Figs.~6(a) and 6(b), are relatively 
small. Specifically, they remain under 0.2 deg. across all scenarios. 
Nevertheless, despite their subtlety, these variations can significantly 
influence the bending mode of water, particularly depending on the prevailing 
temperature and pressure conditions \cite{w-yu20,w-se20}.

\section{Pressure-volume equation of state}

The q-TIP4P/F potential model has been demonstrated to provide accurate
results for the temperature dependence of the molar volume
of water, using path-integral simulations.
Specifically, it predicts a temperature of maximum density of 
280~K, in reasonable agreement with experimental data 
\cite{w-ha09,w-ra10,w-el21}. 
The molar volume derived from PIMD simulations is larger than
that obtained in classical simulations, leading to a concomitant
reduction in water density, as indicated earlier \cite{w-ra10}.
We have verified that our results 
for the temperature dependence of the molar volume,
as well as the isothermal compressibility $\kappa_T$,
at room conditions ($T$ = 300~K, $P$ = 1 bar) agree with those
obtained earlier using PIMD with the q-TIP4P/F interatomic potential
\cite{w-ra10,w-el21}.

\begin{figure}
\vspace{-6mm}
\includegraphics[width=7cm]{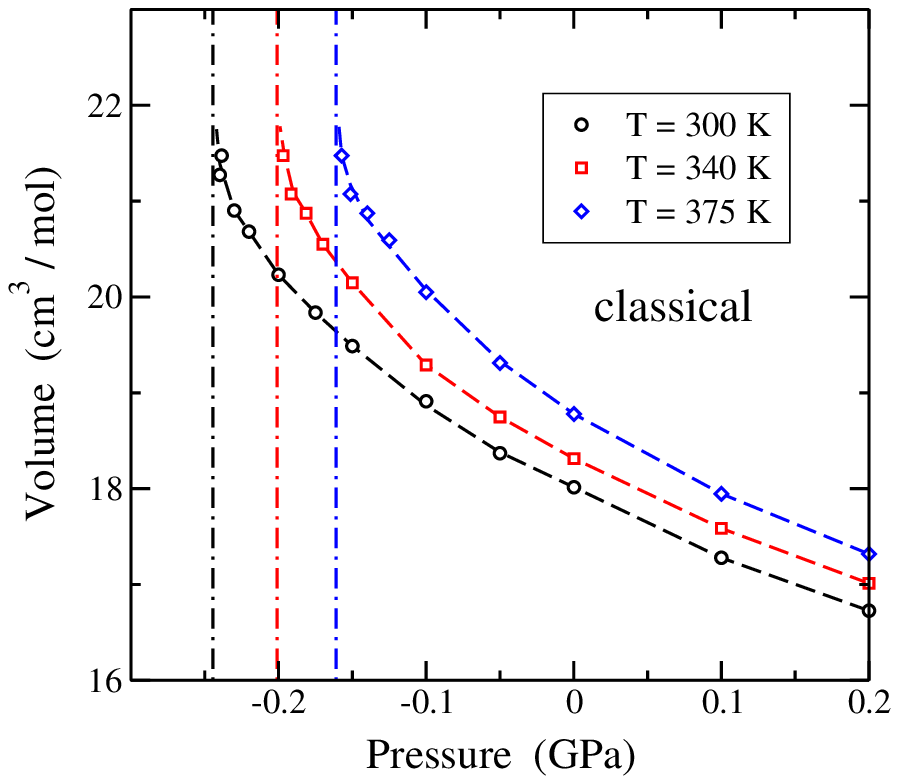}
\vspace{-5mm}
\caption{
Molar volume of water as a function of pressure at three temperatures:
$T$ = 300 K (circles), 340 K (squares), and 375 K (diamonds).
Symbols indicate results of classical MD simulations.
Error bars are on the order of the symbol size.
Dashed lines are guides to the eye.
The vertical dashed-dotted lines indicate the estimated spinodal
pressure for each temperature.
}
\label{f7}
\end{figure}

Under tensile stress, the molar volume of water steadily increases
in the pressure range of mechanical stability.
In Fig.~7, we display a graphical representation illustrating the 
relationship between molar volume and tensile pressure, as determined 
through classical simulations conducted at three distinct temperatures: 
$T$ = 300~K (represented by circles), 340~K (depicted by squares), and 
375~K (indicated by diamonds). The molar volume exhibits an 
accelerated increase with rising tensile pressure, culminating in the 
proximity of the mechanical instability specific to each temperature.

For better clarity, vertical dashed-dotted lines have been incorporated 
in the graph, denoting the spinodal pressure $P_s$ corresponding 
to each temperature. At $P_s$, the rate of change of 
volume with respect to pressure, $(\partial V / \partial P)_T$, 
approaches $-\infty$, signifying a singular point.
The simulations are unable to approach this limit, and the vertical 
dashed-dotted lines serve as markers for this theoretical boundary.

Furthermore, we note that, for each temperature in Fig.~7, 
the three data points in closest proximity to the spinodal pressure 
were obtained from $NVT$ simulations. 
The considerable volume fluctuations observed in $NPT$ simulations 
near the liquid-gas spinodal pressure impede 
an effective sampling of that region of the configuration 
space. This limitation becomes more pronounced at elevated temperatures, 
where the concurrent increase in volume fluctuations increases
the difficulty of accurately exploring the system.
To address this question, we have carried out simulations in 
the canonical ensemble in the segments of the 
configuration space where liquid water maintains a metastable state 
throughout sufficiently extended simulation runs. This 
ensures that the system remains within the metastable phase 
of liquid water for a duration conducive to the precise 
sampling of structural and thermodynamic variables. 
By leveraging $NVT$ simulations in these specific regions, 
we avoid the impact of volume fluctuations, 
allowing for a more reliable and 
comprehensive exploration of the targeted configuration space,
close to the spinodal pressure at the considered temperatures. 

\begin{figure}
\vspace{-6mm}
\includegraphics[width=7cm]{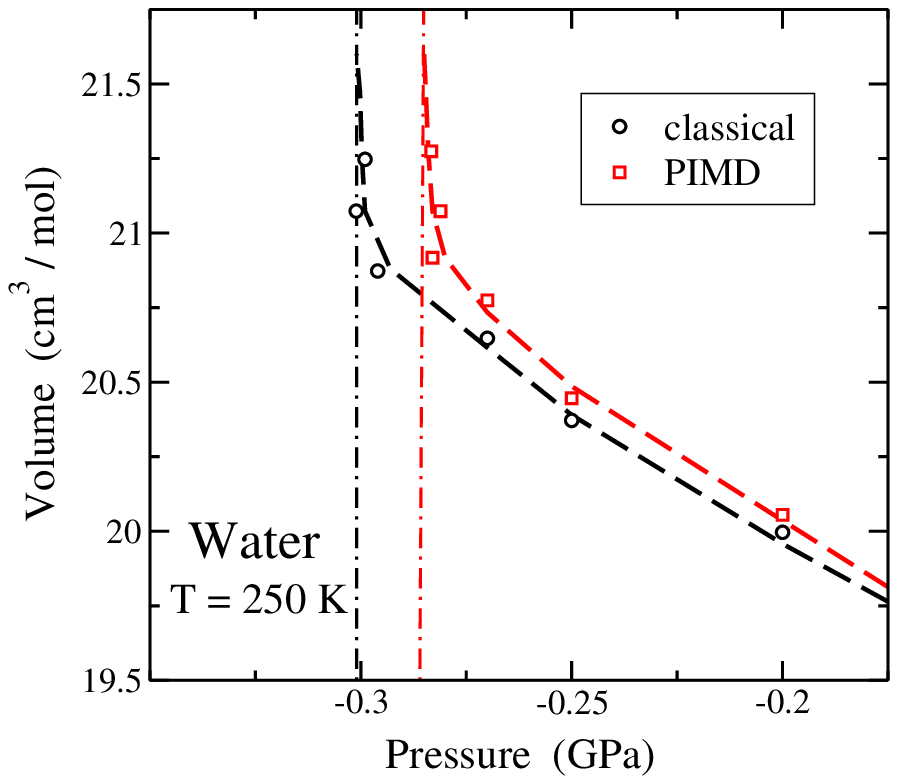}
\vspace{-5mm}
\caption{
Pressure dependence of the molar volume of water at $T = 250$~K.
Symbols represent results of classical MD (circles) and
PIMD simulations (squares).
Error bars are in the order of the symbol size.
Lines are guides to the eye.
Vertical dashed-dotted lines indicate the spinodal pressure
for each approach.
}
\label{f8}
\end{figure}

In Fig.~8, we present the pressure-dependent 
behavior of the molar volume of water, derived from simulations utilizing 
the q-TIP4P/F potential at a temperature of 250~K. 
The symbols in the graph represent simulation outcomes, with circles 
denoting classical MD and squares indicating PIMD simulations.
Within error bars, a coincidence between classical and quantum 
results is observed for $P > -0.1$~GPa, aligning with previous 
findings \cite{w-el21} at the same temperature. 
However, as tensile pressure increases, a noteworthy 
divergence emerges, with quantum simulations yielding volumes surpassing 
their classical counterparts. This deviation intensifies with increasing 
tensile pressure, ultimately culminating in a manifestation of mechanical 
instability at lower tension in the quantum simulations.

From these results we estimate for the liquid-gas spinodal pressure $P_s$
values of $-301$ and $-286$~MPa for the classical and quantum approaches,
respectively. These $P_s$ values are shown in Fig.~8 as 
vertical dashed-dotted lines. We find that nuclear quantum dynamics 
causes a shift in the spinodal pressure of 15~MPa at $T$ = 250~K.
This shift is smaller for higher temperatures, as shown below.

According to the results of our simulations,
this trend persists across a region of temperatures ranging 
from 250 to 375~K. It is noteworthy that while the quantum effect on the 
spinodal pressure is discernible across this temperature range, its 
influence diminishes with rising temperature. At $T$ = 375~K, the quantum 
impact on the spinodal pressure becomes notably subtle, approaching 
the limits of observability in our results. 
This temperature-dependent behavior elucidates the interplay between 
quantum effects, temperature, and mechanical stability, 
providing valuable insights into the thermodynamic properties of water 
under varying conditions.

\section{Spinodal instability}

The liquid expansion caused by tensile stress gives rise to 
a fast increase in the compressibility, which diverges for a 
temperature-dependent pressure $P_s$, where liquid water becomes 
mechanically unstable. 
This is typical of a spinodal point in the $(P, T)$ phase diagram 
\cite{he03b,w-sc95,ra18b,ca85}.
Given a temperature $T$, there is a region of tensile pressure 
where liquid water is metastable, specifically, for $P_s < P < 0$. 
The spinodal line, delineating the unstable phase
($P < P_s$) from the metastable phase, is the locus of points $P_s(T)$
where $\kappa_T$ diverges, or equivalently, where its inverse,
the isothermal bulk modulus, vanishes.
This type of spinodal line has been previously investigated for water 
\cite{w-sp82,w-sp82b,w-ne01,w-ga07,w-az13,w-ca19,w-du20,w-im08d}, as 
well as for various types of solids \cite{w-sc95,w-ma19,he03b}. 
In recent years, research has explored this question 
in two-dimensional materials, where this type of instability 
may also be found under compressive stress \cite{ra18b,ra20}.

Close to a spinodal point, the Helmholtz free energy $F$ at 
temperature $T$ can be written as a Taylor power expansion in 
terms of the volume difference $V_s - V$ \cite{w-sp82,bo94b,ra20}:
\begin{eqnarray}
  F(V, T) & = &  F(V_s(T), T) + \alpha_1(T) \, [V_s(T) - V] +  
	\nonumber  \\
          & + &  \alpha_3(T) \, [V_s(T) - V]^3 + ...  \; .
\label{ffc}
\end{eqnarray}
Here $V_s(T)$ and $F(V_s(T), T)$ are the volume and free energy
at the spinodal point, where one has
$\partial^2 F / \partial V^2 = 0$, and a quadratic term
$[V_s(T) - V]^2$ does not appear in Eq.~(\ref{ffc}):
 $\alpha_2 = 0$.
The coefficients $\alpha_i$ along with the spinodal volume
$V_s$ depend in general on the temperature, although
we will not explicitly write this dependence in the sequel.
The pressure is given by
\begin{equation}
   P = - \frac{\partial F}{\partial V} =
           P_s + 3 \, \alpha_3 \, (V_s - V)^2 + ...   \; ,
\label{pfv}
\end{equation}
and $P_s = \alpha_1$ corresponds to the spinodal pressure and to
the volume $V_s$.

The isothermal compressibility $\kappa_T$ is obtained as
\begin{equation}
\kappa_T = \frac{1}{V} \, \left( \frac{\partial^2 F}{\partial V^2} \right)^{-1} =
   - \frac{1}{V}  \, \left( \frac{\partial P}{\partial V} \right)_T^{-1}   \; ,
\label{bpa}
\end{equation}
and one has for $P$ close to $P_s$:
\begin{equation}
   \kappa_T \sim  (P - P_s)^{-1/2} \; .
\label{ba2}
\end{equation}
The compressibility diverges to infinity for $P \to P_s$, 
as the inverse square root of $P - P_s$.

\begin{figure}
\vspace{-6mm}
\includegraphics[width=7cm]{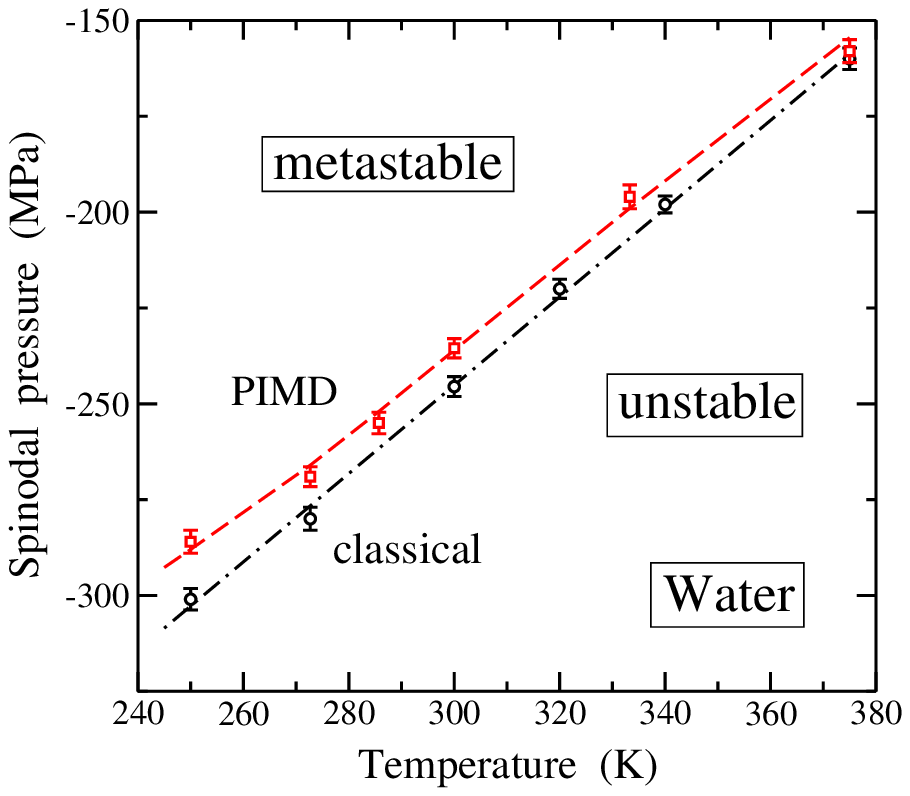}
\vspace{-5mm}
\caption{Temperature dependence of the spinodal pressure, $P_s$,
derived from classical MD (solid circles) and PIMD simulations
(solid squares). The dashed-dotted line is a linear fit to
the classical data. The dashed line through the quantum outcomes
is a guide to the eye.
}
\label{f9}
\end{figure}

In Fig.~9 we display the liquid-gas spinodal pressure $P_s$ of water 
as a function of the temperature, estimated from our classical (circles) 
and PIMD simulations (squares).
We observe a decrease in $P_s$ as the temperature is lowered 
into the region of supercooled water. Our classical results can be 
fitted to a straight line $P_s = b_1 + b_2 T$,
with  $b_1 =  -591$~MPa and $b_2 = $ 1.15 MPa/K
(shown as the dashed-dotted line in Fig.~9). Consequently, we do not 
detect any indication of reentrant behavior (an increase in $P_s$ at 
low temperatures), in agreement to findings in Refs.\cite{w-ne01,w-ga07} 
from classical MD simulations. The anticipated reentrant spinodal, 
extrapolated from various thermodynamic properties of water \cite{w-sp82b}, 
according to the stability-limit conjecture, suggests that the spinodal 
pressure should decrease upon cooling, become negative, and then increase 
after reaching a minimum \cite{w-sp82b,w-az13}.
However, our simulation data are in line with the model proposed by 
Poole {\em et al.} \cite{w-po92}, where the spinodal remains positively 
sloped and extends to larger negative pressures when lowering 
the temperature.

From the outcomes of our PIMD simulations, we observe that 
nuclear quantum dynamics induces a positive shift in the liquid-gas 
spinodal pressure within the considered temperature range. 
Specifically, we find shifts of 10 and 15 MPa at temperatures of 
300 and 250 K, respectively (resulting in $P_s$ becoming less negative). 
Our quantum data appear to deviate slightly from linearity, especially 
at the lowest temperatures depicted in Fig.~9. The quantum effect on 
$P_s$ is found to increase as $T$ is lowered, mirroring trends 
seen in other physical variables calculated from path integral 
simulations.

As indicated in the plot, in each case, classical or quantum,
the liquid exhibits metastability at negative pressures in the region 
above the corresponding dashed-dotted or dashed line. 
Below it, the liquid becomes mechanically unstable, 
leading to a transformation into the gas phase, 
where the volume diverges to infinity under tensile pressure. 
When approaching the line (classical or quantum) from the metastable 
region, the transition may occur well before reaching the spinodal. 
This phenomenon was clearly observed in our isothermal-isobaric 
simulations when increasing the tensile stress near the spinodal.

Our classical results for the liquid-gas spinodal pressure are not far 
from those obtained by Netz {\em et al.} \cite{w-ne01} from MD simulations 
using the SPC/E potential model. Specifically, these authors reported
a value of approximately $-270$~MPa at $T = 250$ K, compared to
our value of $-301$~MPa at the same temperature. 
Gallo {\em et al.} \cite{w-ga07} conducted MD simulations of 
stretched water employing a polarizable potential model, and from
their spinodal pressure results, we interpolate a value of about 
$-330$~MPa at $T$ = 250~K.
The $P_s$ value derived from our classical MD simulations at this
temperature falls intermediate between those obtained using these 
other potential models.

More recently, Biddle {\em et al.} \cite{w-bi17} employed the
TIP4P/2005 interatomic potential to study the thermodynamics of water
in a wide range of temperature and pressure, using classical
MD simulations. From the results presented in their Fig.~1, 
we estimate at $T$ = 250 and 300~K values of 
$P_s \approx -310$ and $-240$~MPa, to be compared with the
outcomes of our classical simulations: $-301$ and $-246$~MPa, 
respectively.  A smaller value for the spinodal pressure at 300~K, 
$P_s \approx -175$~MPa was found in \cite{w-po92}
from an extrapolation of $P-V$ isotherms at negative $P$, derived
by using MD simulations with the ST2 pair potential.

\section{Summary}

In this paper, we have presented the outcomes of PIMD simulations 
for liquid water, under various hydrostatic pressures encompassing 
both compressive and tensile regimes. 
These quantum simulations, carried out at a specific temperature, afford 
a comprehensive analysis of water within pressure ranges where it exhibits 
stability or metastability. Through this methodology, we have quantitatively 
examined several properties of water, with a particular focus on 
delineating its mechanical stability limit.

The significance of quantum effects has been evaluated by contrasting 
the findings derived from PIMD simulations with those emanating from 
classical MD simulations. Specifically, 
we have employed the q-TIP4P/F interatomic potential, which 
has proved to be well-suited for this kind of investigations. 
This potential model adeptly captures numerous structural and thermodynamic 
features across distinct water phases, underscoring its applicability 
and reliability in elucidating the intricate behavior of water molecules.

While the molar volume of water experiences a reduction under increased 
compression at a constant temperature, the interatomic distances exhibit 
a distinct pattern, characteristic of various condensed phases of water. 
Notably, the interatomic distance between oxygen atoms in adjacent water 
molecules decreases with rising compressive pressure, while the 
intramolecular distance $d_{\rm O-H}$ concurrently increases, indicative 
of a simultaneous attenuation in the strength of the covalent bond.

Quantum nuclear motion induces changes in structural variables, 
prompting a detailed analysis of quantum effects on internal and 
kinetic energy, intramolecular distance $d_{\rm O-H}$, bond angle 
$\theta_{\rm H-O-H}$, and molar volume as a function of pressure and 
temperature. Notably, our focus extends to $P < 0$, particularly in 
proximity to the spinodal pressure corresponding to 
each considered temperature.
The influence of quantum nuclear motion manifests in a discernible 
shift of the liquid-gas spinodal pressure, amounting to 15 and 10~MPa 
at temperatures of 250 and 300~K, respectively. Within this temperature 
range, our quantum simulations reveal $P_s$ values spanning from 
$-286$ to $-236$~MPa.

These findings underscore the interplay among temperature, pressure, 
and nuclear quantum motion in shaping the structural dynamics of water 
molecules. This insight not only provides a pathway for further 
exploration but also facilitates a deeper comprehension of the 
nuanced behaviors governing water at the molecular level. 
Such understanding holds potential implications across diverse 
fields, spanning from materials science to environmental studies.

The computational methodology outlined in this paper has demonstrated 
its reliability as a robust tool for elucidating the impact of pressure 
on metastable states within liquids. In particular, it facilitates 
the calculation of the spinodal line under tensile stress as 
a temperature-dependent function. Future investigations in this domain 
are essential to expand upon the results presented herein, especially 
in the context of other liquids. The stability limits of these liquids 
will hinge on the behavior of their compressibility under tensile stress, 
warranting further exploration and analysis.

\begin{acknowledgments}
This work was supported by Ministerio de Ciencia e Innovaci\'on
(Spain) through Grant PID2022-139776NB-C66.
\end{acknowledgments}


\end{document}